\documentclass{article}
\usepackage{amsfonts}
\usepackage{amsmath}
\usepackage{textcomp}
\usepackage{graphicx}

\graphicspath{{converted_graphics/}{/}}

\begin{document}

\title{Alternate derivation of Padmanabhan's differential\\
bulk-surface relation in General Relativity}
\author{Dennis G Smoot \\
Aurora University \\
Center for Teaching and Learning \\
Aurora, IL 60506-4892 \\
dsmoot@aurora.edu}
\date{\today}
\maketitle

\begin{abstract}
A differential bulk-surface relation of the lagrangian of General Relativity
has been derived by Padmanabhan. This has relevance to gravitational
information and degrees of freedom. An alternate derivation is given based
on the differential form gauge theory formulation of gravity due to G\"{o}%
ckeler and Sch\"{u}cker. Also an entropy functional of Padmanabhan and
Paranjape can be rewritten as the G\"{o}ckeler and Sch\"{u}cker lagrangian.
\end{abstract}


\section{Introduction}

Padmanabhan, in the context of standard General Relativity(GR), has derived
a differential relation between the bulk and surface terms of the
Einstein-Hilbert(EH) lagrangian. He has done this in 3 different ways. \cite%
{pad031136}, \cite{padbook}, \cite{pad0608120}, \cite{pad1005}. This
splitting of the GR lagrangian was known historically, but it was
Padmanabhan who found a relation between them. This implies that the bulk
and surface terms, one being the derivative of the other, contain the same
information. According to Padmanabhan this has relevance to both the physics
of black holes and the Universe as a whole. In fact Padmanabhan claims the
gravitational degrees of freedom of a region are on its boundary.

Padmanabhan's general procedure to study relativity is to use accelerated
noninertial Rindler frames the same way Einstein used inertial frames. It
will not be possible to give a detailed exposition of all of Padmanabhan's
results in this short paper but see \cite{pad031136} or \cite{padbook}\ for
a synopsis.

\subsection{The G\"{o}ckeler and Sch\"{u}cker formulation}

The mathematics of this paper is based on the differential form gauge theory
formulation of GR of G\"{o}ckeler and Sch\"{u}cker(G\&S). \cite{GockSchu}
The G\&S action for gravity is%
\begin{equation}
S_{GS}=-\frac{1}{32\pi G}\int\nolimits_{\mathcal{U}}R_{\ \beta }^{\alpha
}\wedge \star (f^{\beta }\wedge f_{\alpha })  \label{gocksch}
\end{equation}

\noindent where $G$ is Newtons gravitational constant, $\mathcal{U}$ is some
open set on a manifold or fiber bundle taken to be $4$-dimensional spacetime
with $\alpha ,\beta =0,\ldots ,3$. The $f$s are an oriented basis of the
cotangent space, $f^{\alpha }=f_{\ \beta }^{\alpha }\,dx^{\beta }$, $%
\mathbb{R}
^{4}$ valued $1$-forms. $\ast $ is the Hodge star, the duality
(complementary) map amongst basis forms. 
\begin{equation}
R=d\omega +\omega \wedge \omega \qquad R^{\alpha \beta }=d\omega ^{\alpha
\beta }+\omega _{~\gamma }^{\alpha }\wedge \omega ^{\gamma \beta }
\end{equation}

\noindent where $\omega $ is a $\mathbf{gl_{4}}$ valued $1$-form. $d$ is the
exterior derivative. and $D$ is the exterior covariant derivative. Thus $R$
is also a $\mathbf{gl_{4}}$ valued $2$-form. (Note: Symbolically $R=D\omega =
$ $d\omega +\omega \wedge \omega $. However $\omega $ does not transform in
the same linear representation of the gauge group as defined by $D$. So $%
R\neq D\omega $) Indices are raised and lowered with the metric $g_{\alpha
\beta }$ whose signature here is $+---$. In Mathematics $\omega $ is called
the connection and in Physics it is called the potential. It is the
generator of parallel translations and is the gauge field. \cite{GockSchu}
In General Relativity the torsion $T=0$. In this G\&S formulation $0\neq
T=Df=df+\omega \wedge f$ and is an $%
\mathbb{R}
^{4}$ valued $2$-form. According to Wald torsion is necessary for spinorial
matter. \cite{Wald84} Also according to Aldrovandi, Barros, and Pereira \cite%
{AldoPer} non$0$ torsion is incompatible with the Principle of Equivalence
because then objects do not move on geodesics. It may be mentioned that G%
\"{o}ckeler and Sch\"{u}cker clarify the gauge structure of GR and\ the
difference between diffeomorphisms and co\"{o}rdinate transformations, often
confused in the literature. \cite[pp 75, 87]{GockSchu}

The G\&S Action, Eqn (\ref{gocksch}), is a gauge theory of GR and reduces to
the EH form under contraction of certain indices. This can be seen as
follows.

GR requires a holonomic frame, namely $f^{\alpha }=dx^{\alpha }$. Otherwise
in a nonholonomic frame $f^{\alpha }=f_{~\zeta }^{\alpha }dx^{\zeta }$ where
the $f_{~\zeta }^{\alpha }\,\epsilon \,%
\mathbb{R}
^{4}$, $\zeta =0,\cdots ,3$. Using the definition of $\ast $, $\ast
(f^{\beta }\wedge f_{\alpha })=\ast (f^{\beta }\wedge f^{\zeta }g_{\zeta
\alpha })\rightarrow $

\noindent $\ast (g_{\zeta \alpha }dx^{\beta }\wedge dx^{\zeta })=\varepsilon
_{\zeta \eta \gamma \delta }g_{\iota \alpha }g^{\iota \zeta }g^{\eta \beta
}dx^{\gamma }\wedge dx^{\delta }\left\vert \det g_{\iota \kappa }\right\vert
^{\frac{1}{2}}$. $\varepsilon $ is the totally antisymmetric $\varepsilon $%
-tensor or Levi-Civita tensor. Then $R_{\ \beta }^{\alpha }\wedge \star
(g_{\zeta \alpha }f^{\beta }\wedge f^{\zeta })\rightarrow -R_{\ \beta
}^{\alpha }\wedge \varepsilon _{\zeta \eta \gamma \delta }g_{\alpha \iota
}g^{\iota \zeta }g^{\eta \beta }(dx^{\gamma }\wedge dx^{\delta })\sqrt{-g}%
=R^{\alpha \beta }\varepsilon _{\alpha \beta \gamma \delta }dx^{\gamma
}\wedge dx^{\delta }\sqrt{-g}$ where $g=\det g_{\iota \kappa }$. Finally
particular co\"{o}rdinates are chosen, i.e. the gauge is fixed.

Now the EH lagrangian requires a scalar curvature so $R^{\alpha \beta }:=%
\frac{1}{2}R_{\;\;\gamma \delta }^{\alpha \beta }dx^{\gamma }\wedge
dx^{\delta }$ is contracted to $R^{\alpha \beta }\rightarrow 2\cdot \frac{1}{%
2}R_{\;\;\alpha \beta }^{\alpha \beta }dx^{\alpha }\wedge dx^{\beta }$. A
factor of $2$ is included because there are $2\times $ less indices.
Substituting into Eqn. (\ref{gocksch}) gives%
\begin{eqnarray}
S_{GS}[\beta ,\omega ] &=&-\frac{1}{32\pi G}\int_{\mathcal{U}}R\varepsilon
_{\alpha \beta \gamma \delta }dx^{\alpha }\wedge dx^{\beta }\wedge
dx^{\gamma }\wedge dx^{\delta }\sqrt{-g}  \notag \\
&=&\frac{1}{32\pi G}\int_{\mathcal{U}}\sqrt{-g}dx^{4}R=S_{EH}[dx,\Gamma ]
\label{gstoeh}
\end{eqnarray}

\noindent where $dx^{\alpha }\wedge dx^{\beta }\wedge dx^{\gamma }\wedge
dx^{\delta }=4!dx^{[\alpha }\otimes dx^{\beta }\otimes dx^{\gamma }\otimes
dx^{\delta ]}$ with the brackets denoting antisymmetrization. So $%
\varepsilon _{\alpha \beta \gamma \delta }dx^{\alpha }\wedge dx^{\beta
}\wedge dx^{\gamma }\wedge dx^{\delta }=\dfrac{1}{4!}4!dx^{[\alpha }\otimes
dx^{\beta }\otimes dx^{\gamma }\otimes dx^{\delta
]}=dx^{0}dx^{1}dx^{2}dx^{3}=d^{4}x$ assuming the latter is oriented. $\Gamma 
$ is the notation for the connection $\omega $ in GR.

The variation $f$ in $S_{GS}[\beta ,\omega ]$ to obtain the Einstein
equations(EE) is done in G\&S

\begin{equation}
G_{\alpha \beta }=8\pi G\tau _{\alpha \beta }
\end{equation}

\noindent where the Einstein tensor $G_{\ \delta }^{\gamma }=R_{\ \ \alpha
\delta }^{\alpha \gamma }-\frac{1}{2}R_{\ \ \alpha \beta }^{\alpha \beta
}\delta _{\ \delta }^{\gamma }$ and the energy-momentum tensor $\tau
_{\alpha \beta }$ is defined as $\tau _{\alpha }=\frac{1}{6}\tau _{\alpha
}^{\ \beta }\varepsilon _{\beta \gamma \delta \zeta }f^{\gamma }\wedge
f^{\delta }\wedge f^{\zeta }$.

Varying $\omega $ Eqn. (\ref{gocksch}) leads to the torsion-spin equations%
\begin{equation}
T^{\alpha }\wedge f^{\beta }\varepsilon _{\gamma \delta \alpha \beta }=-8\pi
GS_{\gamma \delta }
\end{equation}

\noindent where $S_{\gamma \delta }$ is the 3-form spin tensor.

\subsection{The bulk and surface terms}

Substituting $R_{\ \beta }^{\alpha }=d\omega _{\ \beta }^{\alpha }+\omega
_{\ \gamma }^{\alpha }\wedge \omega _{\ \beta }^{\gamma }$ and using the
Hodge star $\ast $ in the G\&S action, Eqn (\ref{gocksch}), gives%
\begin{eqnarray*}
S_{GS}[\beta ,\omega ] &=&-\frac{1}{32\pi G}\left[ \int_{\mathcal{U}}\omega
_{\ \kappa }^{\alpha }\wedge \omega _{~\beta }^{\kappa }\wedge \varepsilon
_{\zeta \eta \gamma \delta }g_{\iota \alpha }g^{\iota \zeta }g^{\eta \beta
}f^{\gamma }\wedge f^{\delta }\sqrt{-g}\right. \\
&&\quad \left. +\int_{\mathcal{U}}d\omega _{\ \beta }^{\alpha }\wedge
\varepsilon _{\zeta \eta \gamma \delta }g_{\iota \alpha }g^{\iota \zeta
}g^{\eta \beta }f^{\gamma }\wedge f^{\delta }\sqrt{-g}\right]
\end{eqnarray*}

Simplifying

\begin{eqnarray}
S_{GS}[\beta ,\omega ] &=&-\frac{1}{32\pi G}\left[ \int_{\mathcal{U}}\omega
_{\ \kappa }^{\alpha }\wedge \omega ^{\kappa \beta }\wedge \varepsilon
_{\alpha \beta \gamma \delta }f^{\gamma }\wedge f^{\delta }\sqrt{-g}\right. 
\notag \\
&&\quad \left. +\int_{\mathcal{U}}d\omega ^{\alpha \beta }\wedge \varepsilon
_{\alpha \beta \gamma \delta }f^{\gamma }\wedge f^{\delta }\sqrt{-g}\right]
\label{split integ}
\end{eqnarray}

\noindent The second term on the right hand side(rhs) can be made an exact
differential in the following manner.%
\begin{eqnarray}
&&d(\omega ^{\alpha \beta }\wedge f^{\gamma }\wedge f^{\delta }\sqrt{-g}%
\varepsilon _{\alpha \beta \gamma \delta })  \notag \\
&=&d(\omega ^{\alpha \beta })\wedge f^{\gamma }\wedge f^{\delta }\sqrt{-g}%
\varepsilon _{\alpha \beta \gamma \delta }-\omega ^{\alpha \beta }\wedge
(df^{\gamma })\wedge f^{\delta }\sqrt{-g}\varepsilon _{\alpha \beta \gamma
\delta }  \notag \\
&&\hspace{0.15in}+\omega ^{\alpha \beta }\wedge f^{\gamma }\wedge
(df^{\delta })\sqrt{-g}\varepsilon _{\alpha \beta \gamma \delta }-\omega
^{\alpha \beta }\wedge f^{\gamma }\wedge f^{\delta }\wedge (d\sqrt{-g}%
)\varepsilon _{\alpha \beta \gamma \delta }  \label{differen}
\end{eqnarray}

\noindent because in any single co\"{o}rdinates $\varepsilon _{\alpha \beta
\gamma \delta }$ being constants, $d\varepsilon _{\alpha \beta \gamma \delta
}=0$ and $\omega $ and $f$ are 1-forms. Note $-\omega ^{\alpha \beta }\wedge
(df^{\gamma })\wedge f^{\delta }\varepsilon _{\alpha \beta \gamma \delta
}=\omega ^{\alpha \beta }\wedge f^{\gamma }\wedge (df^{\delta })\varepsilon
_{\alpha \beta \gamma \delta }$. Substituting (\ref{differen}) in (\ref%
{split integ}) and rearranging gives%
\begin{eqnarray}
S_{GS}[\beta ,\omega ] &=&-\frac{1}{32\pi G}\left[ \int_{\mathcal{U}}\omega
_{\ \zeta }^{\alpha }\wedge \omega ^{\zeta \beta }\wedge f^{\gamma }\wedge
f^{\delta }\sqrt{-g}\varepsilon _{\alpha \beta \gamma \delta }-2\int_{%
\mathcal{U}}\omega ^{\alpha \beta }\wedge f^{\gamma }\wedge (df^{\delta })%
\sqrt{-g}\varepsilon _{\alpha \beta \gamma \delta }\right.  \notag \\
&&\hspace{0.15in}\left. +\omega ^{\alpha \beta }\wedge f^{\gamma }\wedge
f^{\delta }\wedge (d\sqrt{-g})\varepsilon _{\alpha \beta \gamma \delta
}+\int_{\mathcal{U}}d(\omega ^{\alpha \beta }\wedge f^{\gamma }\wedge
f^{\delta }\sqrt{-g}\varepsilon _{\alpha \beta \gamma \delta })\right]
\label{new integ}
\end{eqnarray}

The last term on the rhs is an exact differential and consequently by Stokes
Theorem, $\int_{\mathcal{U}}dA=\int_{\partial \mathcal{U}}A$ with $\partial 
\mathcal{U}$ the boundary of $\mathcal{U}$, is a boundary term. The bulk and
the surface terms become respectively%
\begin{eqnarray}
&&-\frac{1}{32\pi G}\int_{\mathcal{U}}\left\{ \left[ \omega _{\ \zeta
}^{\alpha }\wedge \omega ^{\zeta \beta }\wedge f^{\gamma }\wedge
f^{d}-2\omega ^{\alpha \beta }\wedge f^{\gamma }\wedge (df^{\delta })\right] 
\sqrt{-g}\varepsilon _{\alpha \beta \gamma \delta }\right.  \notag \\
&&\hspace{1in}\left. +\omega ^{\alpha \beta }\wedge f^{\gamma }\wedge
f^{\delta }\wedge (d\sqrt{-g})\varepsilon _{\alpha \beta \gamma \delta
}\right\}  \label{bulk}
\end{eqnarray}

\noindent and

\noindent 
\begin{equation}
-\frac{1}{32\pi G}\int_{\partial \mathcal{U}}\omega ^{\alpha \beta }\wedge
f^{\gamma }\wedge f^{\delta }\sqrt{-g}\varepsilon _{\alpha \beta \gamma
\delta }  \label{surf}
\end{equation}

Note there are $2$ exterior derivatives in the bulk term. $f$ is a $1$-form
so $df$ is a $2$-form and can be assumed to be $df=Cf\wedge f$ with $C$ a(n)
(indexed) scalar function. $\sqrt{-g}=\sqrt{\left\vert \det g_{\alpha \beta
}\right\vert }$ is a scalar function so $d\sqrt{\left\vert \det g_{\alpha
\beta }\right\vert }$ is a $1$-form. Now for the metric $g=g_{\alpha \beta }$
(not $g$ =$\det g_{\alpha \beta }$) $0=Dg=dg-\omega g-\omega ^{T}g$ where $D$
is the exterior covariant derivative, i.e. $dg_{\alpha \beta }=g_{\alpha
\gamma }\omega _{\ \beta }^{\gamma }+\omega _{\ \alpha }^{\gamma }g_{\gamma
\beta }$. This is the metric condition. It is assumed \cite{GockSchu} so
that the metric $g$ is preserved under parallel translation of vectors. This
is also an assumption made in GR but is only a simplifying assumption of
Differential Geometry. It is the assumption that preserves lengths and
angles of vectors under parallel translation. The relation $d\sqrt{-g}=%
\dfrac{1}{2}\sqrt{-g}g^{\alpha \beta }dg_{\alpha \beta }$ is an ordinary
matrix relation. Finally since $\omega $ is a 1-form $\omega =Ff$ with $F$ a 
$\mathbf{gl_{4}}$ function. Thus these relations imply (\ref{bulk}) is free
of exterior derivatives except for (possibly) the ones in the integrand
measure, the $f$s.

If there is a differential relation between the bulk and surface terms, i.e.
if exterior derivative (the differential) of the integrand of the surface
term $=$ the integrand of the bulk term, then%
\begin{equation}
d\omega ^{\alpha \beta }=\omega _{\ \zeta }^{\alpha }\wedge \omega ^{\zeta
\beta }
\end{equation}

\noindent This is because taking the exterior derivative of the surface
integrand Eqn. (\ref{surf}) just undoes the steps in Eqn. (\ref{differen})
and cancels the terms in Eqn (\ref{bulk}).

Then $R=d\omega +\omega \wedge \omega =2\omega \wedge \omega $. This imposes
a condition on $R$. Perhaps more importantly it defines $d\omega $. If $%
R\neq $ $2\omega \wedge \omega $ then there is no differential relation
between the bulk and surface terms. This is true irregardless of whether
assumptions are made for $df$ or $d\sqrt{-g}$. These results follow only
from the G\&S gauge formulation of GR. Again if there is such a differential
bulk-surface relation then the bulk and surface terms contain the same
information, one being the differential of the other.

\subsection{Padmanabhan's differential bulk-surface relation}

Now it is desired to reduce the above to GR to compare with Padmanabhan. It
has already been shown above that $S_{GS}[\beta ,\omega ]$ reduces to $%
S_{EH}[dx,\Gamma ]$ and yield the EE.

Reiterating, for the metric $g_{\alpha \beta }$ to be an isometry \cite%
{GockSchu} $0=Dg=dg-g\omega -\omega ^{T}g$. A connection $\omega $ which
satisfies this condition is called metric and was assumed by Einstein in GR.
In co\"{o}rdinates $dg_{\alpha \beta }=g_{\alpha \gamma }\omega _{\ \beta
}^{\gamma }+\omega _{\ \alpha }^{\gamma }g_{\gamma \beta }=\omega _{\alpha
\beta }+\omega _{\beta \alpha }$. Also $d\sqrt{-g}=\dfrac{1}{2}\sqrt{-g}%
g^{\alpha \beta }dg_{\alpha \beta }$. So $d\sqrt{-g}=\omega _{\ \zeta
}^{\zeta }\sqrt{-g}$. Further in GR $0=T=Df=df+\omega \wedge f$, also
assumed by Einstein. Thus $df^{\alpha }=-\omega _{\ \beta }^{\alpha }\wedge
f^{\beta }$.

Substituting for $d\sqrt{-g}$ and $df$ from the previous $2$ lines into (\ref%
{new integ}) gives

\begin{eqnarray}
S_{GS}[\beta ,\omega ] &=&-\frac{1}{32\pi G}\left[ \int_{\mathcal{U}}\omega
_{\ \zeta }^{\alpha }\wedge \omega ^{\zeta \beta }\wedge f^{\gamma }\wedge
f^{\delta }\sqrt{-g}\varepsilon _{\alpha \beta \gamma \delta }+2\int_{%
\mathcal{U}}\omega ^{\alpha \beta }\wedge f^{\gamma }\wedge \omega _{~\zeta
}^{\delta }\wedge f^{\zeta }\sqrt{-g}\varepsilon _{\alpha \beta \gamma
\delta }\right.  \notag \\
&&\qquad \left. +\omega ^{\alpha \beta }\wedge f^{\gamma }\wedge f^{\delta
}\wedge \omega _{\ \zeta }^{\zeta }\sqrt{-g}\varepsilon _{\alpha \beta
\gamma \delta }+\int_{\mathcal{U}}d(\omega ^{\alpha \beta }\wedge f^{\gamma
}\wedge f^{\delta }\sqrt{-g}\varepsilon _{\alpha \beta \gamma \delta })%
\right]
\end{eqnarray}

Applying Stokes Theorem and setting the $1$-form $\omega _{\ \gamma
}^{\alpha }=\omega _{\ \gamma \zeta }^{\alpha }f^{\zeta }$ gives

\begin{eqnarray}
S_{GS}[\beta ,\omega ] &=&-\frac{1}{32\pi G}\left[ \int_{\mathcal{U}}\omega
_{\ \zeta \eta }^{\alpha }\omega _{\hspace{0.05in}\ \ \kappa }^{\zeta \beta
}f^{\eta }\wedge f^{\kappa }\wedge f^{\gamma }\wedge f^{\delta }\sqrt{-g}%
\varepsilon _{\alpha \beta \gamma \delta }\right.  \notag \\
&&\quad -2\int_{\mathcal{U}}\omega _{~~~\eta }^{\alpha \beta }\omega
_{~\zeta \kappa }^{\delta }f^{\eta }\wedge f^{\kappa }\wedge f^{\gamma
}\wedge f^{\zeta }\sqrt{-g}\varepsilon _{\alpha \beta \gamma \delta }  \notag
\\
&&\quad +\omega _{~~~\eta }^{\alpha \beta }\omega _{\hspace{0.05in}\ \zeta
\kappa }^{\zeta }f^{\eta }\wedge f^{\kappa }\wedge f^{\gamma }\wedge
f^{\delta }\varepsilon _{\alpha \beta \gamma \delta }\sqrt{-g}  \notag \\
&&\quad \left. +\int_{\partial \mathcal{U}}\omega _{\hspace{0.09in}\zeta
}^{\alpha \beta }f^{\zeta }\wedge f^{\gamma }\wedge f^{\delta }\sqrt{-g}%
\varepsilon _{\alpha \beta \gamma \delta }\right]  \label{final}
\end{eqnarray}

This expression embodies the assumptions of GR except the integrands are not
contracted to scalar functions like in GR. (Recall $R=d\omega +\omega \wedge
\omega $, $R^{ab}=\frac{1}{2}R_{\;\;\gamma \delta }^{\alpha \beta
}f^{r}\wedge f^{s}$, and $R_{\;\;\gamma \delta }^{\alpha \beta }\rightarrow
R_{\;\;\alpha \beta }^{\alpha \beta }$.) Also the basis of the cotangent
space in GR should be holonomic, $f\rightarrow dx$.

These integrands will reduce under contraction to Padmanabhan's bulk and
surface Lagrangians. The expressions of Padmanabhan's bulk and surface terms
(neglecting the integrand measures) are \cite{pad031136}%
\begin{equation}
\text{bulk: }\sqrt{-g}g^{ik}\left( \Gamma _{\hspace{0.06in}il}^{m}\Gamma _{%
\hspace{0.02in}km}^{l}-\Gamma _{\hspace{0.02in}ik}^{l}\Gamma _{\hspace{0.07in%
}lm}^{m}\right) \text{ \ surface: }-\sqrt{-g}\left( g^{ck}\Gamma _{\hspace{%
0.07in}km}^{m}-g^{ik}\Gamma _{\hspace{0.03in}ik}^{c}\right)  \label{padbs}
\end{equation}

\noindent Padmanabhan's latin indice\noindent s which run $c,i,k,l,m=0,...,3$
have been retained.

Getting the terms to match requires some trial and error. For the first term
on the rhs of Eqn. (\ref{final}), $\eta \rightarrow \beta $ and $\kappa
\rightarrow \alpha $. Then $\omega _{\ \zeta \eta }^{\alpha }\omega _{%
\hspace{0.05in}\ \ \kappa }^{\zeta \beta }\overset{\eta \rightarrow \beta }{%
\underset{\kappa \rightarrow \alpha }{\longrightarrow }}\omega _{\ \zeta
\beta }^{\alpha }\omega _{\hspace{0.05in}\ \ \alpha }^{\zeta \beta }=\omega
_{\hspace{0.05in}\ \ \alpha }^{\zeta \beta }\omega _{\ \zeta \beta }^{\alpha
}\rightarrow g^{ik}\omega _{\hspace{0.07in}il}^{m}\omega _{\hspace{0.02in}%
mk}^{l}$ is the same as Padmanabhan's first bulk term in (\ref{padbs}), if $%
\omega _{\ \zeta \eta }^{\alpha }$ is $\Gamma _{jk}^{i}$. Also it was
necessary to assume $\omega $ is symmetric in its lower indices as is the
Christoffel symbol $\Gamma $. The contractions make the integrand measure or
volume form be $-f^{\beta }\wedge f^{\alpha }\wedge f^{\gamma }\wedge
f^{\delta }\sqrt{-g}\varepsilon _{\alpha \beta \gamma \delta }$. This agees
with the conventional EH action as in Padmanabhan as the outside sign is
taken to be $+$ and not $-$ as here in G\&S. The other terms are done almost
the same. For the second term on the rhs of (\ref{final}) $\omega _{~~~\eta
}^{\alpha \beta }\omega _{~\zeta \kappa }^{\delta }\overset{\eta \rightarrow
\beta }{\underset{\kappa \rightarrow \alpha ,\,\zeta \rightarrow \delta }{%
\longrightarrow }}\omega _{~~~\beta }^{\alpha \beta }\wedge \omega _{~\delta
\alpha }^{\delta }\rightarrow g^{ik}\omega _{\ ik}^{l}\omega _{\ \delta
\alpha }^{\delta }$. In the third term on the rhs of (\ref{final}) again as
before $\eta \rightarrow \beta ,$ $\kappa \rightarrow \alpha ,\,$\ and $%
\zeta \rightarrow \delta $. The same term results as in the previous case
with $-2$ of the previous and $+1$ of the last giving the second bulk term.
The volume forms all become $-f^{\beta }\wedge f^{\alpha }\wedge f^{\gamma
}\wedge f^{\delta }\sqrt{-g}\varepsilon _{\alpha \beta \gamma \delta }$.

In the boundary term in (\ref{final}) $\omega _{\hspace{0.09in}\zeta
}^{\alpha \beta }$ must be contracted over both $\alpha $ and $\beta $.
Writing this out%
\begin{eqnarray}
&&\int_{\partial \mathcal{U}}\omega _{\hspace{0.09in}\zeta }^{\alpha \beta
}f^{\zeta }\wedge f^{\gamma }\wedge f^{\delta }\sqrt{-g}\varepsilon _{\alpha
\beta \gamma \delta }  \notag \\
&=&\int_{\partial \mathcal{U}}\omega _{\hspace{0.09in}\zeta }^{\alpha \zeta
}f^{\zeta }\wedge f^{\gamma }\wedge f^{\delta }\sqrt{-g}\varepsilon _{\alpha
\zeta \gamma \delta }+\int_{\partial \mathcal{U}}\omega _{\hspace{0.09in}%
\zeta }^{\zeta \beta }f^{\zeta }\wedge f^{\gamma }\wedge f^{\delta }\sqrt{-g}%
\varepsilon _{\zeta \beta \gamma \delta }  \notag \\
&=&\int_{\partial \mathcal{U}}\left( \omega _{\hspace{0.09in}\zeta }^{\alpha
\zeta }-\omega _{\hspace{0.09in}i}^{i\alpha }\right) f^{\zeta }\wedge
f^{\gamma }\wedge f^{\delta }\sqrt{-g}\varepsilon _{\alpha \zeta \gamma
\delta }
\end{eqnarray}

\noindent which is the surface term when account is taken of the overall $-$
in Eqn. (\ref{final}).

The following diagram summarizes the results.\footnote{%
This diagram is due to Louis Kauffman.}

\begin{figure}[h]
\centering
\includegraphics[width=5.67in,height=2.47in,keepaspectratio]{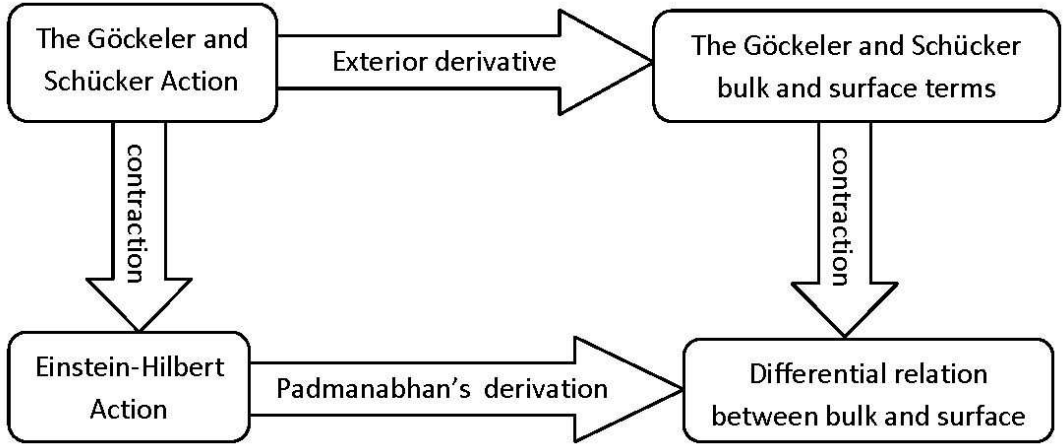}
\caption{Commutative diagram summarizing above derivation.}
\label{fig:commdiag}
\end{figure}

\subsection{Implications of the results}

The above calculations imply $d\omega =\omega \wedge \omega $ so that $%
R=2\omega \wedge \omega $ or if preferred $R=2d\omega $. What may be more
significant however is that $d\omega =\omega \wedge \omega $ defines $%
d\omega $.

To try to understand such results recall GR requires the torsion $T=0$. Then 
$0=T=df+\omega \wedge f$ implies $df=-\omega \wedge f$. So $-\omega \wedge
f=df=\frac{1}{2}Cf\wedge f$, where the assumption $df^{\alpha }=\frac{1}{2}%
C_{\hspace{0.04in}\beta \gamma }^{\alpha }f^{\beta }\wedge f^{\gamma }$ has
been made because $df$ is a $2$-form, with $C_{\hspace{0.04in}\beta \gamma
}^{\alpha }$ antisymmetric in $\beta $ and $\gamma $. Then $\left( \frac{1}{2%
}Cf-\omega \right) \wedge f=0$, and by tensor properties of $\wedge $, $%
\omega =\frac{1}{2}Cf$. Thus $R=2d\omega =Cdf$.

Conversely suppose there is a differential relation between the bulk and
surface terms so that $R=2\omega \wedge \omega $ but the torsion is present
and arbitrary. This is outside the scope of GR. Then $\omega \wedge T=\omega
\wedge df+\omega \wedge \omega \wedge f=\omega \wedge df+\dfrac{1}{2}R\wedge
f$. This is a constraint on the torsion $T$. If $T=0$ as in GR then $R\wedge
f$ $=-2\omega \wedge df$ is a another condition on $R$. Given the
limitations of time it has not been possible to develop this further or to
represent these conclusions in GR.

\section{The entropy functional of Padmanabhan and Paranjape}

In \cite{pad0701003} Padmanabhan and Paranjape postulate an entropy
functional based on elasticity. Gravity as elasticity is Sakharovs paradigm.
For co\"{o}rdinates $x^{a}$, the elastic deformation vector $\xi ^{a}$ with $%
x^{a}\rightarrow x^{a}+\xi ^{a}(x)$, describes the elastic displacement of a
solid. In this section latin indices $a,b=0,\ldots ,3$ will be used
throughout. The quadratic elastic functional is applied to spacetime and is
postulated by them to be%
\begin{equation}
S[\xi ^{\alpha }]=\int_{\mathcal{V}}d^{D}x\sqrt{-g}\left( 4P_{cd}^{\ \
ab}\nabla _{a}\xi ^{c}\nabla _{b}\xi ^{d}-T_{ab}\xi ^{a}\xi ^{b}\right)
\label{elastic}
\end{equation}%
In elasticity the first term on the rhs would correspond to the quadratic
displacement field with $P_{cd}^{\ \ ab}$ constant. The $\xi ^{a}$ are now
being taken as spacetime vectors. The second term corresponds to an
energy-matter term and will be ignored. From this action principle
Padmanabhan and Paranjape can derive the Einstein equations in $4$ and
higher dimensions as well as higher order corrections using the
Lanczos-Lovelock lagrangians \cite{pad0608120}. Further Padmanabhan and
Paranjape are able with this action to derive Wald's Entropy \cite{Waldent}.

Padmanabhan and Paranjape show the first term in the integrand in (\ref%
{elastic})%
\begin{equation}
\mathcal{I}=4P_{cd}^{\ \ ab}\nabla _{a}\xi ^{c}\nabla _{b}\xi ^{d}%
\boldsymbol{\epsilon }
\end{equation}

\noindent where $\boldsymbol{\epsilon }=\dfrac{1}{D!}\epsilon _{a_{1}\cdots
a_{D}}\boldsymbol{\omega }^{a_{1}}\wedge \cdots \wedge \boldsymbol{\omega }%
^{a_{D}}$, $\boldsymbol{\omega }^{i}=\mathbf{d}x^{i}$, $\left( \nabla
_{b}\xi ^{a}\right) \omega ^{b}=\mathbf{d}\xi ^{a}+\boldsymbol{\omega }%
_{~b}^{a}\xi ^{b}$, $\boldsymbol{\omega }_{~b}^{a}=\Gamma _{bc}^{a}%
\boldsymbol{\omega }^{c}$ can be written as%
\begin{equation}
\mathcal{I}=4\ast \mathbf{P}_{ab}\wedge \left( \mathbf{d}\boldsymbol{\xi }%
\right) ^{a}\wedge \left( \mathbf{d}\boldsymbol{\xi }\right) ^{b}
\end{equation}

\noindent where $\mathbf{P}^{ab}=\dfrac{1}{2!}P_{\hspace{0.09in}cd}^{ab}%
\boldsymbol{\omega }^{c}\wedge \boldsymbol{\omega }^{d}$, $\boldsymbol{\xi }%
=\xi ^{a}\mathbf{e}_{a}$, $\mathbf{e}_{a}$ basis vectors, $\left( \mathbf{d}%
\boldsymbol{\xi }\right) ^{a}=\left( \nabla _{b}\xi ^{a}\right) \boldsymbol{%
\omega }^{b}$ and $\ast \mathbf{P}_{ab}=\dfrac{1}{(D-2)!}\dfrac{1}{2!}%
P_{cd}^{\hspace{0.1in}ab}\epsilon _{cda_{1}\cdots a_{D-2}}\boldsymbol{\omega 
}^{a_{1}}\wedge \cdots \wedge \boldsymbol{\omega }^{a_{D-2}}$. (The boldface
type is just following Padmanabhan and Paranjape's conventions.) The $%
\epsilon $ here is the permutation symbol. The $\sqrt{-g}$ is kept in the
integrand measure and not used in the Hodge star $\ast $. Thus (ignoring the
matter term) $S[\xi ]$ becomes 
\begin{equation}
S[\xi ]=4\int_{\mathcal{V}}\ast \mathbf{P}_{ab}\wedge \left( \mathbf{d}%
\boldsymbol{\xi }\right) ^{a}\wedge \left( \mathbf{d}\boldsymbol{\xi }%
\right) ^{b}
\end{equation}

It will now be shown that this Action can be rewritten as a G\"{o}ckeler and
Sch\"{u}cker action $S_{GS}[\beta ,\omega ]$ of the previous section.

Writing the integrand out in full%
\begin{eqnarray}
&&\ast \mathbf{P}_{ab}\wedge \left( \mathbf{d}\boldsymbol{\xi }\right)
^{a}\wedge \left( \mathbf{d}\boldsymbol{\xi }\right) ^{b}  \notag \\
&=&\dfrac{1}{(D-2)!}\dfrac{1}{2!}P_{ab}^{\hspace{0.1in}cd}\epsilon
_{cda_{1}\cdots a_{D-2}}\boldsymbol{\omega }^{a_{1}}\wedge \cdots \wedge 
\boldsymbol{\omega }^{a_{D-2}}\wedge \boldsymbol{\beta }^{a}\wedge 
\boldsymbol{\beta }^{b}  \label{P&P integr}
\end{eqnarray}

\noindent with $\left( \mathbf{d}\boldsymbol{\xi }\right) ^{a}$ being a $1$%
-form set $=\boldsymbol{\beta }^{a}$. Note these $\boldsymbol{\beta }$s are
not holonomic.

On the other hand in $D$ dimensions the integrand in the G\"{o}ckeler and Sch%
\"{u}cker action $S_{GS}[\beta ,\omega ]$ is%
\begin{eqnarray}
&&-R_{~b}^{a}\wedge \ast (\beta ^{b}\wedge \beta _{a})=-R_{ab}\wedge \ast
(\beta ^{b}\wedge \beta ^{a})=R_{ab}\wedge \ast (\beta ^{a}\wedge \beta ^{b})
\notag \\
&=&\dfrac{1}{(D-2)!}\dfrac{1}{2!}R_{abcd}\beta ^{c}\wedge \beta ^{d}\epsilon
_{aba_{1}\cdots a_{D-2}}\sqrt{-g}g^{aa}g^{bb}\beta ^{a_{1}}\wedge \cdots
\wedge \beta ^{a_{D-2}}
\end{eqnarray}

\noindent where $R_{ab}=\dfrac{1}{2!}R_{abcd}\beta ^{c}\wedge \beta ^{d}$.
The $\beta $s are not necessarily holonomic\noindent .

Interchanging the indices $a,b$ and $c,d$ in the previous line%
\begin{eqnarray}
&=&\dfrac{1}{(D-2)!}\dfrac{1}{2!}R_{\hspace{0.1in}ab}^{cd}\beta ^{a}\wedge
\beta ^{b}\sqrt{-g}\epsilon _{cda_{1}\cdots a_{D-2}}\beta ^{a_{1}}\wedge
\cdots \wedge \beta ^{a_{D-2}}  \notag \\
&=&\dfrac{1}{(D-2)!}\dfrac{1}{2!}R_{\hspace{0.1in}ab}^{cd}\epsilon
_{cda_{1}\cdots a_{D-2}}\beta ^{a_{1}}\wedge \cdots \wedge \beta
^{a_{D-2}}\wedge \beta ^{a}\wedge \beta ^{b}
\end{eqnarray}

\noindent which is the same as Eqn. (\ref{P&P integr}) if $R_{\hspace{0.1in}%
ab}^{cd}=P_{ab}^{\hspace{0.1in}cd}$ and the $\beta ^{a_{i}}$s are chosen to
match the $\boldsymbol{\omega }$s. The $\sqrt{-g}$ has been absorbed into
the $\epsilon $ permutation symbol and the name retained. Thus Padmanabhan
and Paranjape's elasticity formulation can be reexpressed in terms of the
curvature. It may be mentioned that the Padmanabhan and Paranjape's paper 
\cite{pad0701003} is quite elegant.

Note that if this reformulation is accepted it implies that Wald's entropy
can also be derived from the G\&S formulation. Also it would imply the gauge
invariance of G\&S GR\ would apply in some sense to the Wald entropy. Work
on this is ongoing.

\end{document}